\documentclass[1p]{elsarticle}

\usepackage{lineno,hyperref}
\usepackage{amsmath}
\usepackage{subcaption}
\usepackage{xcolor}
\usepackage{commath}
\modulolinenumbers[5]

\makeatletter
\providecommand{\doi}[1]{%
  \begingroup
    \let\bibinfo\@secondoftwo
    \urlstyle{rm}%
    \href{http://dx.doi.org/#1}{%
      doi:\discretionary{}{}{}%
      \nolinkurl{#1}%
    }%
  \endgroup
}
\makeatother

\journal{Comptes Rendus Physique}
\begin{document}

\begin{frontmatter}

\title{Low-temperature marginal ferromagnetism explains anomalous scale-free correlations in natural flocks}

\author[isc]{Andrea Cavagna}
\author[isc,sap]{Antonio Culla\corref{cor}}
\ead{a.culla@uniroma1.com}
\cortext[cor]{Corresponding author}
\author[isc,sap]{Luca Di Carlo} 
\author[isc,sap,infn]{Irene Giardina} 
\author[IstLiqBio,conicet,DepFis]{Tomas S. Grigera}

\address[isc]{
Istituto Sistemi Complessi, Consiglio Nazionale delle Ricerche,00185 Roma, Italy}
\address[sap]{Dipartimento di Fisica, Universit\'a Sapienza,00185 Roma, Italy}
\address[infn]{INFN, Unit\`a di Roma 1, 00185 Roma, Italy}
\address[IstLiqBio]{Instituto de F\'isica de L\'iquidos y Sistemas Biol\'ogicos (IFLYSIB), CONICET y Universidad Nacional de La Plata, Calle 59 no. 789, B1900BTE La Plata, Argentina}
\address[conicet]{CCT CONICET La Plata, Consejo Nacional de Investigaciones Cient\'ificas y T\'ecnicas (CONICET), Argentina}
\address[DepFis]{Departamento de F\'isica, Facultad de Ciencias Exactas, Universidad Nacional de La Plata, Argentina}

\begin{abstract}
We introduce a new ferromagnetic model capable of reproducing one of the most intriguing properties of collective behaviour in starling flocks, namely the fact that strong collective order  coexists with scale-free correlations of the modulus of the microscopic degrees of freedom, that is the birds' speeds. The key idea of the new theory is that the single-particle potential needed to bound the modulus of the microscopic degrees of freedom around a finite value, is marginal, that is, it has zero curvature. We study the model by using mean-field approximation and Monte Carlo simulations in three dimensions, complemented by finite-size scaling analysis. While at the standard critical temperature, $T_c$, the properties of the marginal model are exactly the same as a normal ferromagnet with continuous symmetry-breaking, our results show that a novel zero-temperature critical point emerges, so that in its deeply ordered phase the marginal model develops divergent susceptibility and correlation length of the modulus of the microscopic degrees of freedom, in complete analogy with experimental data on natural flocks of starlings.
\end{abstract}


\end{frontmatter}

\section{Introduction}

\begin{figure}[]
\centering
\begin{subfigure}{0.5\textwidth}
\includegraphics[scale=0.16]{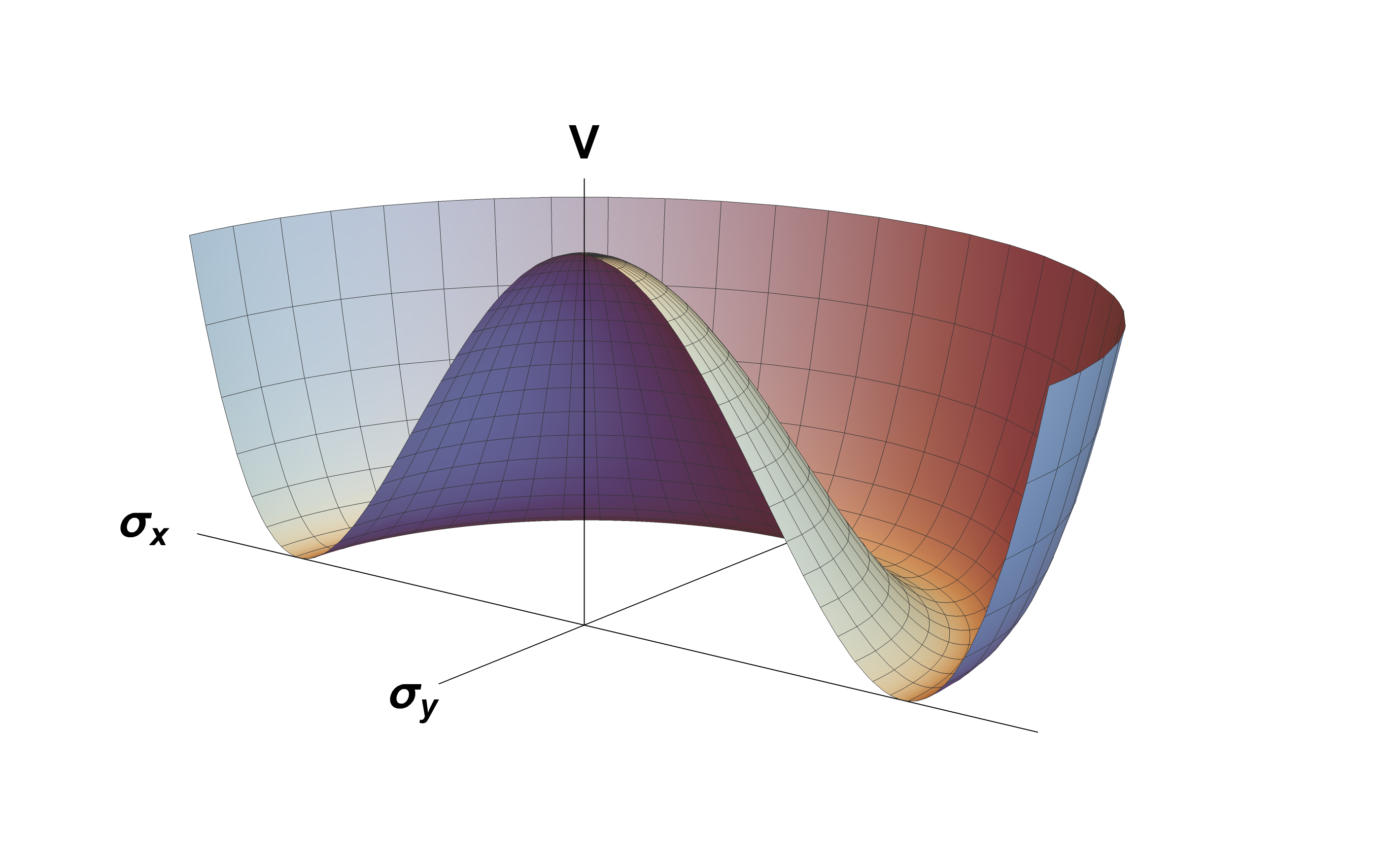}
\end{subfigure}%
\begin{subfigure}{0.5\textwidth}
\includegraphics[scale=0.165]{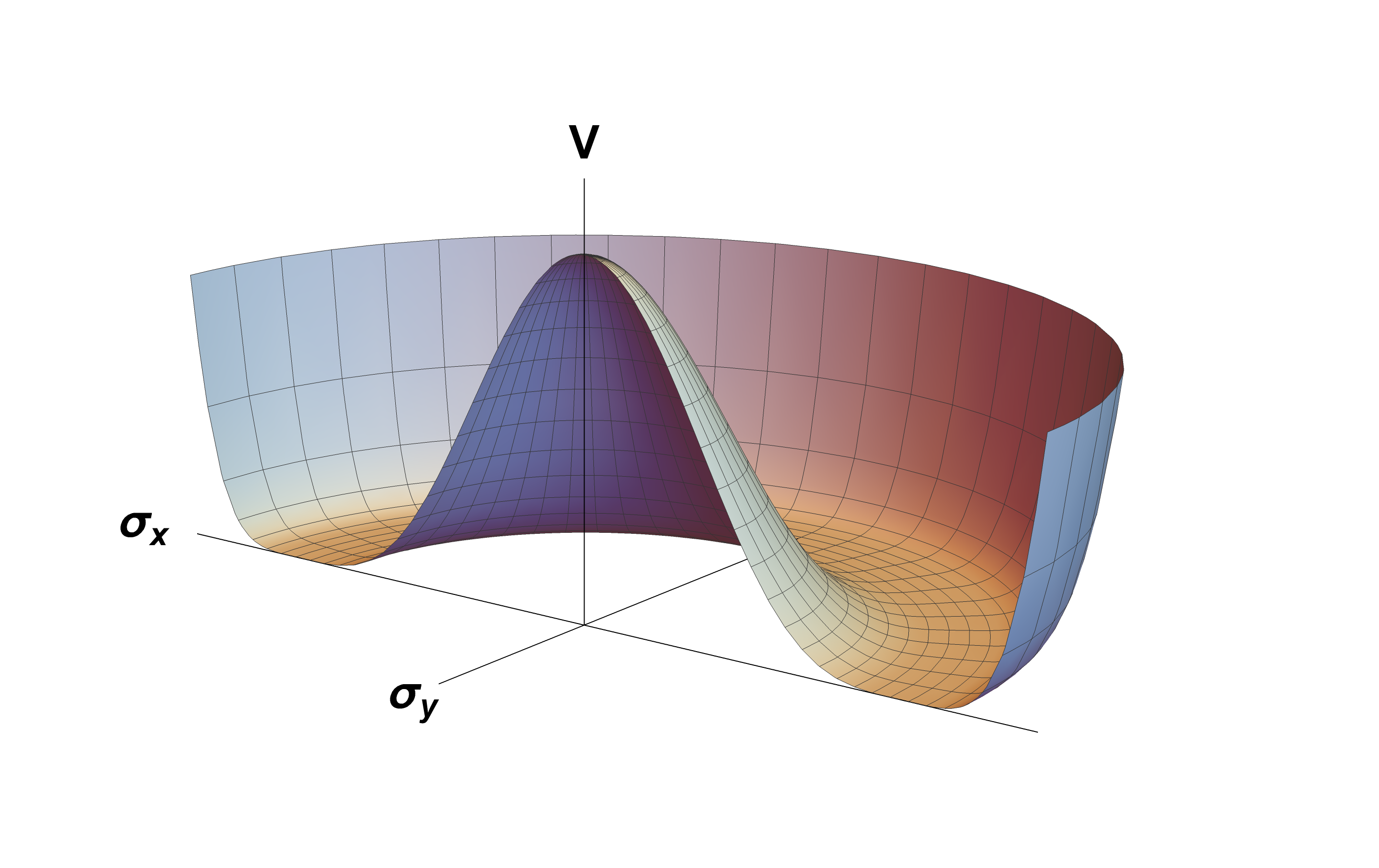}
\end{subfigure}
\caption{Left: In the standard $\mathrm O(n)$ potential (here represented for $n=2$), the marginal modes are the $(n-1)$ Goldstone's transverse modes, while the curvature orthogonal to that is non-zero. Right: In the marginal potential, on the other hand, there is an extra marginal mode in the longitudinal direction.}
\label{twopot}
\end{figure}

Ferromagnetic models with $\mathrm O(n)$ rotational symmetry in their low temperature phase develop a nonzero order parameter and massless Goldstone modes, which are consequence of the spontaneously broken continuous symmetry \cite{goldstone1961field, goldstone1962broken}. In turn, Goldstone modes give rise to infinite susceptibility and correlation length in the whole sym\-me\-try-broken phase \cite{patashinskii_book}. In a finite-size system, the fact that the bulk correlation length is infinite has the practical consequence that the spatial range of the correlation function scales with the system's size, $L$; in other words, there is no intrinsic length scale in the system apart from $L$ itself, a physical state of affairs one normally describes by saying that the system has scale-free correlations \cite{privman1990finite}. In contrast with the emergence of massless (or zero, or marginal) Goldstone modes, in a normal  $\mathrm O(n)$ ferromagnet the {\it modulus} of the microscopic degrees of freedom remains a massive mode, hence it does {\it not} develop scale-free correlations in the ordered phase \cite{ryder1996quantum}. If we use the classic ``Mexican hat" potential as a reference framework (Fig. \ref{twopot}, left), the Goldstone mode is represented by the flat direction circling around the potential (and therefore enabling the $\mathrm O(n)$ rotational symmetry), while the modulus mode is represented by the orthogonal direction, which is steep, namely non-flat, and whose non-zero curvature gives rise to a finite correlation length.

Natural flocks of starlings represent a notable exception to this physical scenario. Bird flocks are systems with obvious $\mathrm O(n)$ symmetry\footnote{This is true as long as we are considering the phenomenon of arial display, namely flocks performing dynamical evolutions on top of their roosting site, and strictly on the plane orthogonal to gravity, which is a clear symmetry-breaking factor; in the case of migrating flocks, where an external preferential direction exists, the symmetry is lost.} 
and a very large order parameter (or polarization), due to the tendency of each individual to align to its neighbours \cite{vicsek_review}. Indeed flocking models based on a self-propelled generalization of low temperature ferromagnets (whose most notable archetype is the Vicsek model \cite{vicsek+al_95}), have been quite successful in describing collective animal behaviour \cite{chate+al_08, cavagna+al_15}. The overarching idea of all these models is that ferromagnetism is essentially a mechanism for {\it imitation}: the role of the magnetic spins is played by the birds' velocities, and the ferromagnetic interaction grants a mutual matching of the neighbours' velocities and speed. In this context, the existence of a Goldstone mode, with consequent scale-free nature of the spatial correlations, has been actually observed in real experiments: the spatial span of the velocity correlation function, which measures the size of the regions over which velocity fluctuations are correlated, scales with the linear size $L$ of the system \cite{cavagna+al_10}. However, another thing that experiments show is that, at variance with normal $\mathrm O(n)$ ferromagnetic systems, natural flocks have scale-free correlations {\it also of the  modulus} of the velocities, namely of the birds' speed \cite{cavagna+al_10}. This is extremely odd, as in the deeply ordered phase, in which flocks live, one would expect the potential regulating the bird's speed to be steep, hence giving a short-range correlation function of the modulus. Hence, flocks' phenomenology suggests that some further marginal mode, beyond the one prescribed by the Goldstone's theorem, is regulating the birds' velocities. Reconciling the paradox  of strong collective order coexisting with scale-free modulus correlations is the task of this work.

A significant step towards solving this problem has been done in \cite{bialek+al_14PNAS}, where it was used a maximum entropy approach to infer directly from the data a model able to reproduce quite accurately the speed correlations. Essentially, the key ingredient of the model in \cite{bialek+al_14PNAS} is a Gaussian term regulating the fluctuations of the speed.
What the inference shows is that, in order to give rise to the correct scale-free correlations of the speed, the coupling constant $g$ of this speed-regulating Gaussian term had to be small; more precisely, within the Gaussian theory there is a simple relation showing that the smaller the speed control $g$ is, the larger the correlation length of the speed; if $g$ is so small that the correlation length gets larger than the system's size, the resulting speed correlations are scale-free. The theory developed in  \cite{bialek+al_14PNAS} has, however, a problematic aspect: being the model Gaussian, the same coupling constant $g$ that needs to be small in order to make modulus fluctuations scale-free, is also the only bound for the absolute value of the velocities; this means that in the thermodynamic limit, $L\to\infty$, in order to have scale-free correlations one needs $g\to 0$, so that the Hamiltonian is actually unbounded. Hence, the calculation of \cite{bialek+al_14PNAS} can only be interpreted as a Gaussian expansion of a more complete, yet unknown, theory. This is the theory we want to develop here.

The central idea of our approach simply emerges from the comparison of the two panels in Fig. \ref{twopot}. In a standard $\mathrm O(n)$ model, there is a single-particle bare potential that keeps the modulus of the microscopic degrees of freedom bounded to some finite reference value; in this way one enforces a soft constraint on the microscopic degrees of freedom, which is an alternative to the hard constraint imposed, for example, by the classical Heisenberg model and also by the Vicsek model \cite{vicsek+al_95}. Of course, thanks to the universality guaranteed by the renormalization group, soft vs hard constraints makes no difference at all to determine the critical properties of the system, which are only ruled by symmetry and dimensions \cite{binney_book}. At low temperatures the system breaks the $\mathrm O(n)$ symmetry and starts fluctuating around one of the (continuously) many ground states of the potential. Although the transverse fluctuations are boosted by the Goldstone mechanism, the fluctuations of the modulus of the microscopic variables are suppressed by the presence of a non-zero longitudinal second derivative of the potential; this is why the modulus is not a scale-free mode in normal $\mathrm O(n)$ models. 

This scenario, however, also suggests a possible solution to our problem. We can use a potential that has the same $\mathrm O(n)$ symmetry, but it also has {\it zero} second derivative along the longitudinal direction (Fig. \ref{twopot}, right).  At very low temperatures, where energy dominates over entropy, the potential will dominate the fluctuations, including those of the modulus, which therefore should become {\it marginal}, namely scale-free. If this is true, we also expect that by raising the temperature, not only we decrease the order parameter, but we certainly add entropic fluctuations to the theory, which should therefore build an effective mass to the modulus fluctuations, thus lowering their correlation length. Finally, at even larger temperatures, entropy should take over, and the system should reach the standard, finite-temperature, critical point, where marginality of all components of the order parameter is reached again. This -- admittedly rather speculative -- scenario is the one we test in the present work.

Before we provide the details of our calculation, a remark is in order. We will develop an {\it equilibrium} theory, namely the interaction network we will consider  will be fixed, rather than time-dependent as in self-propelled systems, which flocks most definitely are. We do this for two reasons. First, there is solid evidence indicating that, at least for natural flocks of starlings, the time needed for the interaction network to significantly rearrange is far larger that the time needed for the birds' velocities to locally relax, so that a quasi-equilibrium approximation makes sense \cite{mora2016local}. Second, even though the precise underlying theory of flocks is surely self-propelled and out of equilibrium, it is possible that the explanation of scale-free speed correlations may actually depend on some sub-part of the theory which is not crucially dependent on self-propulsion, as the single particle potential regulating the speed; if an equilibrium explanation works, then its self-propelled counterpart is likely to work too. Of course, we cannot rule out {\it a priori} the hypothesis that the true explanation of scale-free speed correlations is in fact an intrinsic off-equilibrium feature of the system, not requiring any particular speed-fixing potential. Once such alternative will be presented, only the comparison with the experiments will be able to select the right theory.

\section{The marginal model}

We consider a class of ferromagnetic models with $\mathrm O(n)$ symmetry, defined by the Hamiltonian \cite{Ma_book},
\begin{equation}
H = \frac{J}{2} \sum_{ij}^N n_{ij} (\boldsymbol{\sigma}_i - \boldsymbol{\sigma}_j)^2 +  \sum_i^N V(\boldsymbol\sigma_i\cdot\boldsymbol\sigma_i) - \mathbf{h} \cdot \sum_i^N \boldsymbol{\sigma}_i \ ,
\label{Ham}
\end{equation}
where the $N$ microscopic degrees of freedom $\boldsymbol{\sigma}_i$ are vectors of dimension $n$, representing the spins in the ferromagnetic context, or the velocities in the context of flocking models. The first term is a standard ferromagnetic (i.e. alignment) interaction, of strength $J$, where $n_{ij}$ represents the interaction (or adjacency) matrix, namely $n_{ij}=1$ if the sites $i$ and $j$ interact with each other, $n_{ij}=0$ if they do not; in our simulation $n_{ij}$ will be nonzero only for $i$ and $j$ nearest neighbours. The discrete space structure is a simple cubic lattice of side $L$, in $d = 3$, and $n = 3$, which implies $N=L^3$. The external field $\mathbf{h}$ is coupled to $\boldsymbol{\sigma}_i$, in order to calculate various quantities in what follows. The single-particle potential $V$ is needed to keep the modulus of each vector $\boldsymbol{\sigma}_i$ (which we indicate as $\sigma_i$) close to some reference nonzero value, giving rise to the competition between energy and entropy necessary to produce a second order phase transition \cite{goldenfeld_lectures_1992}. The model has a Gibbs-Boltzmann equilibrium probability distribution, $P \sim \exp{(-\beta H)}$, where $\beta$ is the inverse of the temperature $T$, which is a measure of the noise in the system, related to the accuracy of each bird in monitoring its neighbours' velocity and regulating its own. 

The single-particle potential $V$ is chosen to be a function of $\boldsymbol\sigma_i\cdot\boldsymbol\sigma_i$, hence sharing the same $\mathrm O(n)$ symmetry as the ferromagnetic interaction. In this way, 
the whole low temperature, ordered phase of the model is characterised by the spontaneous breaking of the continuous $\mathrm O(n)$ symmetry, which, through Goldstone's theorem, provides scale-free correlations of the standard susceptibility and correlation length \cite{goldstone1961field, goldstone1962broken}. As we have seen, the challenge is to find a model capable of developing also scale-free correlations of the modulus in the deeply ordered phase. We consider the two following potentials (see Fig. \ref{twopot}), 
\begin{align}
V(\boldsymbol\sigma\cdot\boldsymbol\sigma) &= \lambda \; (\boldsymbol\sigma\cdot\boldsymbol\sigma -1)^2 \quad , \quad {\rm{standard}} \quad 
\\
 \quad V(\boldsymbol\sigma\cdot\boldsymbol\sigma) &= \lambda \; (\boldsymbol\sigma\cdot\boldsymbol\sigma -1)^4 \quad , \quad \rm{marginal} \ ,
\end{align}
where $\lambda$ is a parameter with the same physical dimensions as $J$ and $T$.
The standard potential (Fig. \ref{twopot}, left) is the one giving rise to the classic Heisenberg and $\mathrm O(n)$ ferromagnetic phenomenology \cite{patashinskii_book}; it does {\it not} have scale-free correlations of the modulus when in the ordered phase, and it will be studied by us chiefly as a reference case, to provide a comparison with our new model. The {\it marginal} potential (Fig. \ref{twopot}, right), on the other hand, has the simplest algebraic form necessary to provide a zero second derivative not only along the transverse direction (Goldstone mode), but also along the longitudinal direction, and it is our candidate to produce scale-free correlations of the modulus in the deeply ordered phase. The idea is the following: the magnetic susceptibility is given by the inverse of the second derivative of the Gibbs free energy in its minimum, namely in the equilibrium state. If the system is at very low temperature, the entropic contribution to the Gibbs free energy is sub-leading, so that, in the vicinity of its minimum, the free energy is not that far from the bare energy; because at very low temperature alignment will be high, the ferromagnetic term is negligible, so that the Gibbs free energy (close to its minimum) will be similar to the single particle potential $V$. Hence, it seems reasonable that the vanishing longitudinal second derivative of the marginal potential might cause a divergence of the modulus susceptibility.  Let us first check this idea at mean-field level.


\section{Mean-field model}

In the mean-field case the nearest-neighbour interaction is replaced by a term that links every spin with each other \cite{parisi_book}, thus we have the fully-connected Hamiltonian,
\begin{equation}
H = \frac{J}{2N} \sum_{ij} (\boldsymbol{\sigma}_i - \boldsymbol{\sigma}_j)^2 + \lambda \sum_i (\boldsymbol\sigma_i\cdot\boldsymbol\sigma_i -1)^4 \ ,
\label{HamMF}
\end{equation}
where the coupling $J$ has been divided by the number of sites in the system $N$ to have an extensive Hamiltonian. The external field will not be needed here, so it has been set to $0$. To simplify the notation, in what follows we will indicate any $\mathrm O(n)$-symmetric function $f(\boldsymbol\sigma\cdot\boldsymbol\sigma)$, as $f(\sigma)$. The partition function of the system $Z$ is:
\begin{equation}
Z = \int \prod_k \dif \boldsymbol{\sigma}_k \ e^{- \beta H} = \int \dif \mathbf{m} \int \prod_k \dif \boldsymbol{\sigma}_k \ e^{- \beta H} \ \delta \left(\beta N\mathbf{m} - \beta \sum_j \boldsymbol{\sigma}_j\right) \ ,
\end{equation}
where we have explicitly introduced the magnetization $\mathbf m= (1/N) \sum_i \boldsymbol \sigma_i$ and we have disregarded all subleading factors of order $(\log N)/N$ or higher. By using a Fourier representation of the $\delta$-function, we obtain, 
\begin{equation}
Z = 
\int \dif \mathbf{m} \ e^{N\beta J m^2}
\int d\mathbf x \ \exp\left\{\beta N\left[
\mathbf m \cdot\mathbf x + \frac{1}{\beta}\log\int \dif \boldsymbol{\sigma} \ e^{-\beta (J \sigma^2+ V(\sigma)+\boldsymbol\sigma\cdot\mathbf x)}
\right]
\right\}  \ ,
\label{partfunc}
\end{equation}
where $m=|\mathbf m|$ and $\sigma = | \boldsymbol \sigma|$.
In the limit $N\to\infty$ we can use the saddle-point method to evaluate the integral in $\mathbf x$, which gives the equation fixing the saddle-point value $\mathbf x_0$ as 
a function of $\mathbf m$ (and $\beta$), 
\begin{equation}
\mathbf{m} = \frac{\int \dif \boldsymbol{\sigma} \ \boldsymbol\sigma \ e^{- \beta\left[ S(\sigma)+\mathbf x_0\cdot\boldsymbol\sigma\right]}}{\int \dif \boldsymbol{\sigma} \ 
e^{- \beta\left[ S(\sigma)+\mathbf x_0\cdot\boldsymbol\sigma\right]}
}
\label{mvalue}
\end{equation}
with, $S(\sigma) = J \sigma^2 + V(\sigma)$. We note that, in the saddle point, the magnetization is equal to its equilibrium value, $\mathbf m_\mathrm{eq}= (1/N) \sum_i \left<\boldsymbol \sigma_i \right>$.
We finally obtain,
\begin{equation}
Z = \int \dif \mathbf{m} \ e^{-\beta N g(m)}  \ ,
\end{equation}
where the mean-field Gibbs free energy per particle, $g(m)$, is therefore given by,
\begin{equation}
g(m) = -Jm^2 - \mathbf{m} \cdot \mathbf{x}_0({\mathbf m}) - \frac{1}{\beta} \ln \int \dif  \boldsymbol{\sigma} \ 
e^{- \beta\left[ S(\sigma)+\mathbf x_0(m)\cdot\boldsymbol\sigma\right]} \ .
\label{gofm}
\end{equation}
Notice that $g(m)$ is a $\mathrm O(n)$-symmetric function of the modulus of the magnetization. The mean-field Gibbs free energy is linked to the probability of $m$ by the relation, $P(m)= e^{-\beta N g(m)}/Z$. 

\begin{figure}[]
\centering
\begin{subfigure}{0.5\textwidth}
\includegraphics[scale=0.27]{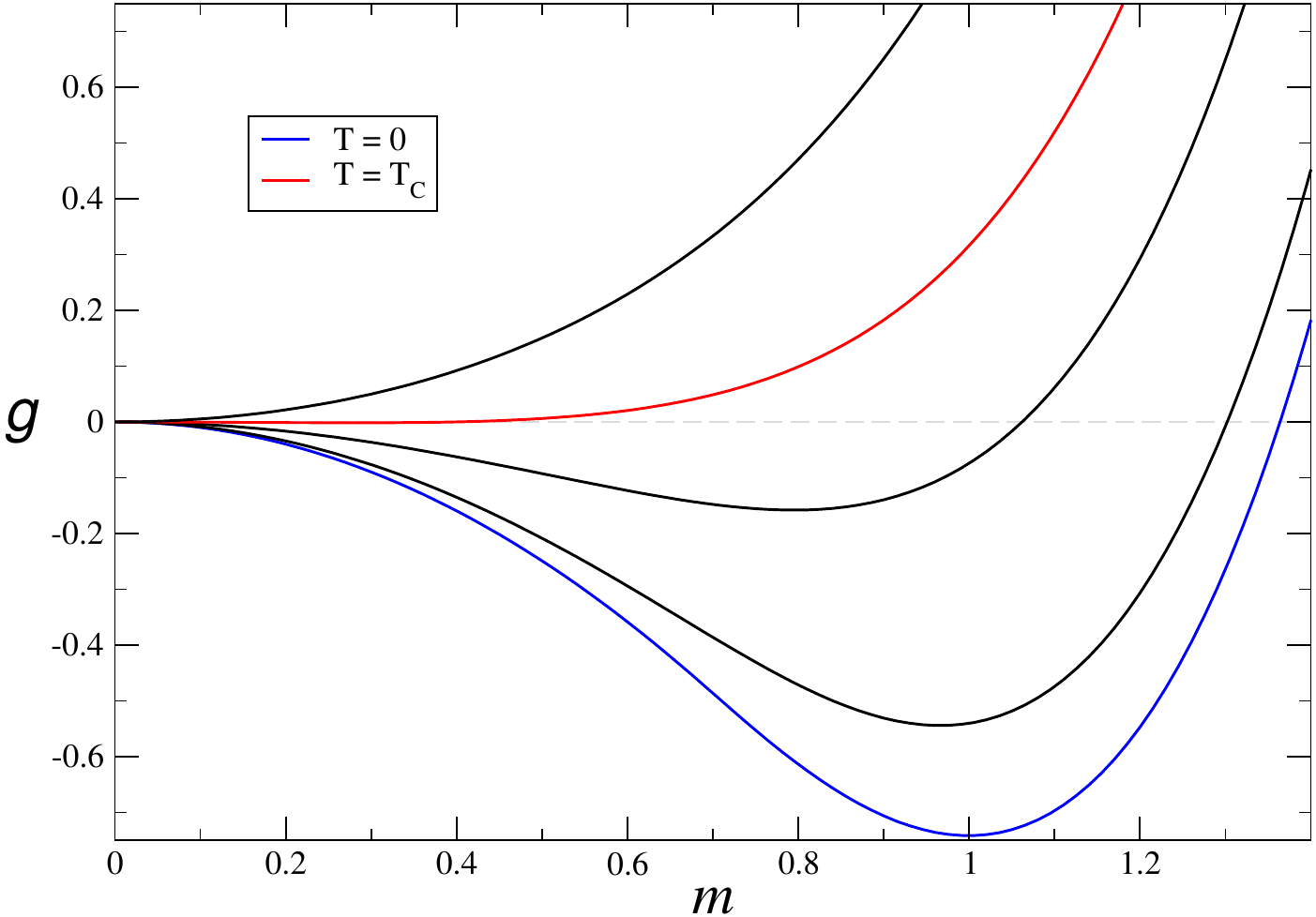}
\end{subfigure}%
\begin{subfigure}{0.5\textwidth}
\includegraphics[scale=0.27]{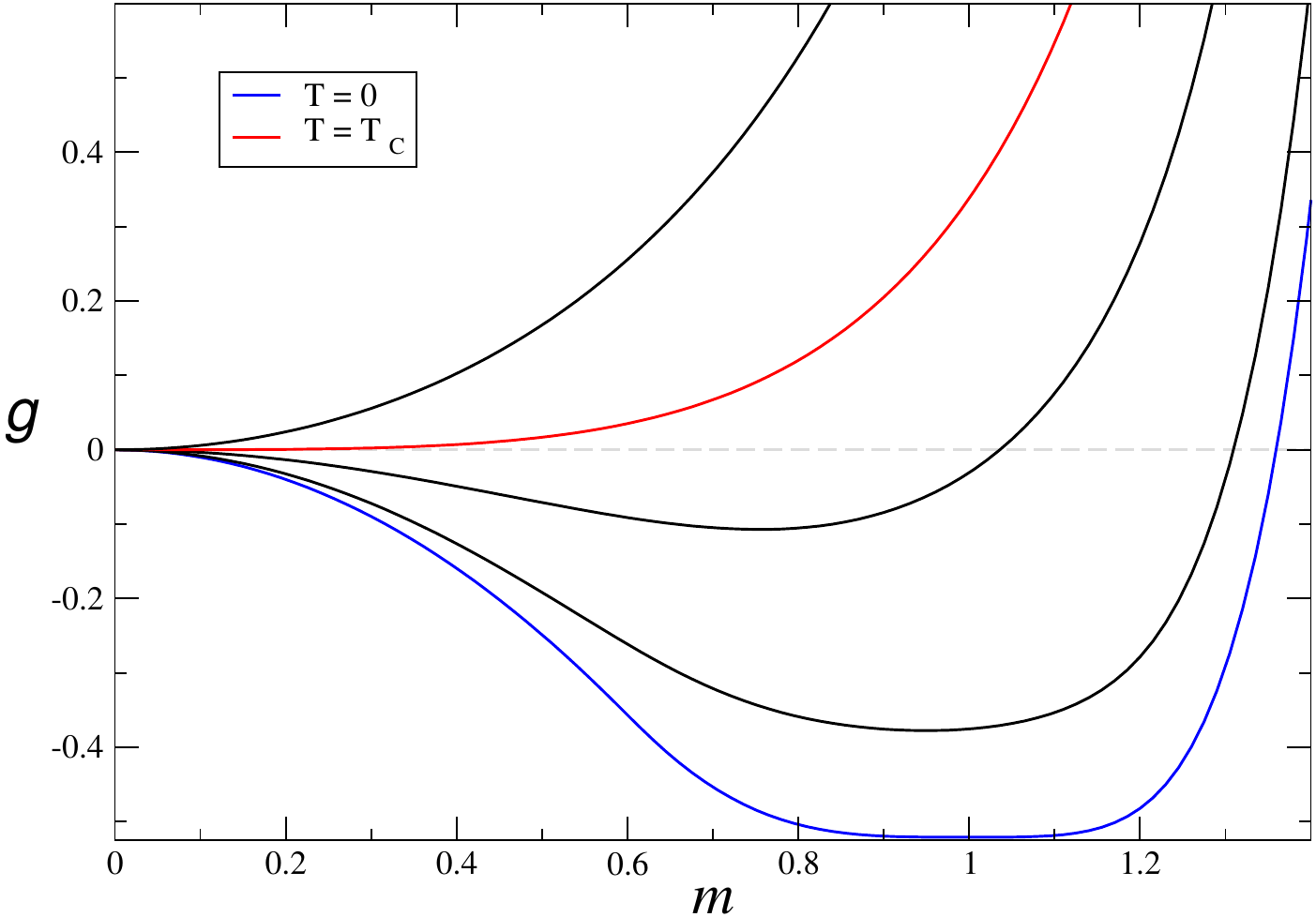}
\end{subfigure}
\caption{Mean-field Gibbs free energy, $g(m)$, for $n=3$, at different temperatures. Left: In the standard $\mathrm O(n)$ model, the curvature of the free energy in its minimum vanishes only at the finite critical temperature, $T_c$. Right: In the marginal model, the free energy has a zero-curvature minimum also at $T=0$, giving rise to the divergence of the modulus susceptibility. By raising the temperature the minimum acquires a finite second derivative, thanks to fluctuations, so that a nonzero mass emerges. By further increasing the temperature, the standard critical point $T_c$ is reached, where the mass of the whole vectorial degree of freedom $\mathbf m$ is zero again, as the curvature of $g(m)$ in $m=0$. The curvature of the free energy in the equilibrium state reaches a maximum some way between $T=0$ and $T=T_c$.}
\label{meanfieldfig}
\end{figure}

From \eqref{gofm}, we can compute $g(m)$ numerically for the marginal potential; the results are shown in Fig. \ref{meanfieldfig} (right) for $n=3$. As we expected, at $T=0$ the free-energy has a flat (i.e. marginal) minimum, which, as we shall soon see, is responsible for a zero-temperature divergent susceptibility of the modulus. This marginal minimum is of course the zero-temperature relic of the minimum of the marginal potential. The interesting point is that by raising the temperature, the free energy minimum develops a nonzero curvature (namely a mass) due to the entropic effects, hence lowering the modulus susceptibility. By further increasing the temperature, $g(m)$ becomes flat again at the finite (bare) critical temperature, $T_c$, as it happens for any standard ferromagnetic model. At, $T_c$, though, the order parameter (i.e. the abscissa of the minimum) goes to zero, hence it makes no sense to distinguish between modulus and direction, and one has just one (normal) diverging susceptibility. Finally, above $T_c$ the paramagnetic phase takes over.  


As expected, the novel behaviour of the marginal model emerges for low temperature. Hence, we perform a saddle-point expansion for $\beta \gg 1$ in both equations
\eqref{mvalue} and \eqref{gofm}. The algebra is rather intricate in the case of the marginal potential, because the first nonzero derivative of $V(\sigma)$ in its saddle point
$\sigma=1$ is the fourth, rather than the second as in the standard potential. We first calculate the equilibrium value of the magnetization, namely the minimum of $g(m)$,
\begin{equation}
m_\mathrm{eq} \simeq 1 - \frac{V^{'''''}(1)}{8V^{''''}(1)}\frac{T}{J} = 1 - \frac{5}{4}\frac{T}{J} \label{magncorr} \ .
\end{equation}
where we gave the exact form of the coefficients for $n=1$, although the scaling with $T$ and $J$ does not depend on $n$.
The susceptibility of the modulus of the magnetization is given by ($x_0=|{\mathbf x}_0|$),
\begin{equation}
\chi_\mathrm{mod}^{-1} = g''(m_\mathrm{eq}) = -2J - \eval{\dod{x_0}{m}}_{m=m_\mathrm{eq}} \ ,
\end{equation}
which, to leading order in $T$, gives,
\begin{equation}
\chi_\mathrm{mod} \simeq \frac{4}{V^{''''}(1)}\frac{J}{T} = \frac{1}{90}\frac{J}{\lambda T} \ .
\label{gamma}
\end{equation}
We conclude that, at mean field level, the marginal potential gives a standard critical transition at a finite temperature $T_c$ and a novel zero-temperature divergence of
the susceptibility of the modulus of the order parameter, with mean-field critical exponent,
\begin{equation}
\gamma_\mathrm{mod}=1  \quad \quad \mathrm{mean \ field}  \ .
\end{equation}
The mean field results go exactly in the expected direction and are therefore encouraging. It remains  to be seen 
whether this behaviour is preserved in finite dimension. To do this, we will perform numerical simulations.


\section{Monte Carlo simulations}
To study the properties of the marginal model, we perform Monte Carlo (MC) simulations on a $d=3$ cubic lattice. We used the Metropolis algorithm \cite{barkema2001monte}, applied to continuous spins: at each trial a random quantity uniformly distributed between $ \pm \Delta$ is added to each component of the selected spin. The value of $\Delta$ is chosen heuristically for each temperature by fixing the acceptance rate to $\sim 0.23$ \cite{roberts1997}. Simulations were made for different sizes, from $L=8$ up to $L=20$. In all simulations we set $J=1$, $\lambda=1$. The spins are three dimensional, $n=3$. All thermal averages, $\langle\cdot\rangle$ are MC time averages.

\begin{figure}[]
\centering
\begin{subfigure}{0.5\textwidth}
\includegraphics[scale=0.27]{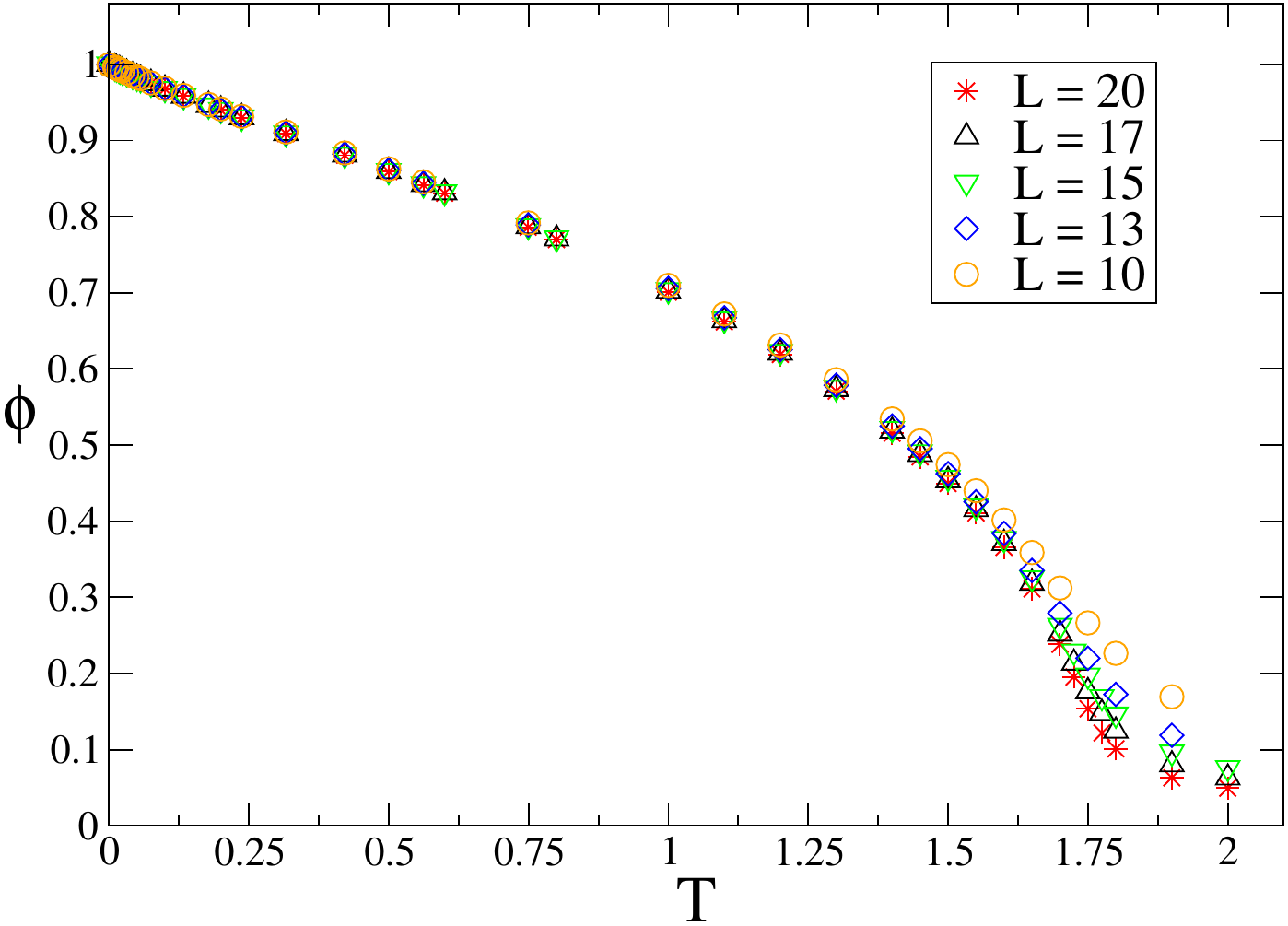}
\caption{}
\end{subfigure}%
\begin{subfigure}{0.5\textwidth}
\includegraphics[scale=0.27]{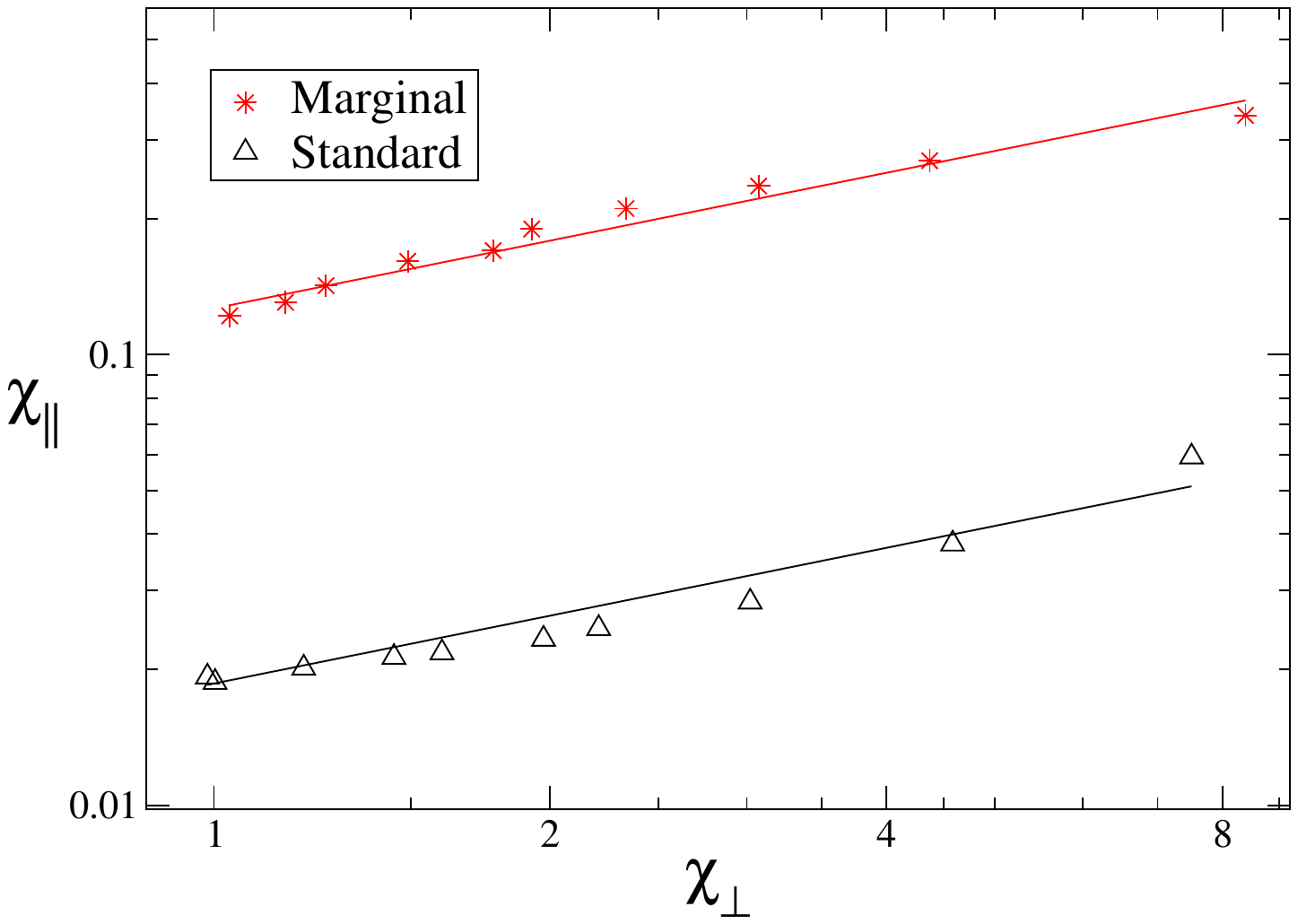}
\caption{}
\end{subfigure}
\caption{(a) Marginal model: order parameter (polarization) as a function of the temperature for several sizes. As can be seen from the plot, the system undergoes a phase transition from a polarized phase for small temperatures, to a disordered phase for high temperatures. The critical temperature of the system is around $T \simeq 1.7$. Notice that the polarization goes to $1$ linearly for $T\to 0$. (b) Longitudinal susceptibility as a function of the transverse one, parametric plot in magnetic field (log-log scale). For both models the relations between susceptibilities, equation \eqref{PPmagic}, holds in the symmetry-broken phase; in $d=3$ we have $\chi_{\parallel} (h) \sim \chi_{\perp} (h)^{1/2}$, so lines are best fit to the data  with exponent fixed to $1/2$. System parameters: size $L = 10$, temperature $T = 0.5$, field $h = (h, 0, 0)$ with $h$ varying from $0.1$ to $1$.}
\label{polsusc}
\end{figure}

\subsection{Tests of the marginal model}

For $T$ smaller but close to $T_c$, namely in proximity of the critical fixed point, the marginal model must have the same  behaviour as the standard $\mathrm{O}(n)$ models. First of all, we check that a standard ordering transition occurs at some finite $T_c$. To this aim we compute the polarization order parameter\footnote{Even though the polarization $\phi$ is not the actual magnetization modulus $m$, it has the advantage of being unaffected by the wandering of spins in finite-size MC simulations, thus it can be measured and it provides reliable information about the magnitude of the system's magnetization. Indeed, $\phi$ differs from the theoretical value of $m$ only by a small offset, that goes to zero with $T$.},
\begin{equation}
 \phi =  \left< \left| \frac{1}{N} \sum_i {\boldsymbol{\sigma}_i}  \right| \right> \ .
 \label{polar}
\end{equation}
The order parameter of the marginal model for different sizes is shown in Fig. \ref{polsusc}a , from which we see that a standard ordering transition occurs around $T_c\sim 1.7$, and that the order parameter goes linearly to $1$ at zero-temperature.

In order to test the behaviour of the marginal model in the symmetry-broken phase, $T<T_c$ (but in the proximity of $T_c$), we calculate the 
transverse and the longitudinal susceptibilities, in the presence of an external field 
$\mathbf h$. Since the system is at equilibrium, the susceptibility matrix is given by \cite{parisi_book},
\begin{equation}
\chi_{\alpha \gamma} = \frac{d\langle\sigma^{(\alpha)}\rangle}{dh^{(\gamma)}}
= \frac{\beta}{N} \sum_{i,j} \left<\sigma_i^{(\alpha)}\sigma_j^{(\gamma)}\right>-\left<\sigma_i^{(\alpha)}\right>\left<\sigma_j^{(\gamma)}\right>,
 \label{suscfdt}
\end{equation}
where $\alpha, \gamma = x, y, z$. In the symmetry-broken phase, fluctuations are dominated by their transverse components (with respect to the order parameter direction) \cite{goldenfeld_lectures_1992}, hence the susceptibility matrix is dominated by $\chi_{\perp}(h)$, where 
$h=|\mathbf h|$. For $h\to 0$ (and $T<T_c$), Goldstone's theorem prescribes that $\chi_\perp(h) \to\infty$: this is the standard
Goldstone marginal mode, responsible for scale-free correlations of the transverse fluctuations in the ordered phase. The less intuitive, and yet classic, result is that also the {\it longitudinal} susceptibility, $\chi_\parallel(h)$, diverges for $h\to 0$ \cite{patashinskii_book}. This is a consequence of the coupling of the longitudinal fluctuations with the transverse ones; essentially, if we call $\theta$ the (small) phase fluctuation of the spins, transverse fluctuations are proportional to $\sin\theta\sim \theta$, whereas longitudinal fluctuations are proportional to $1-\cos\theta\sim \theta^2$. This implies that, in the limit $h\to 0$, standard 
$\mathrm O(n)$ models satisfy the relation \cite{pata1973chi_longi,BrezinWallace1973longi},
\begin{equation}
\chi_{\parallel} (h) \sim \chi_{\perp} (h)^{\epsilon/2}  \ .
\label{PPmagic}
\end{equation}
where $\epsilon=4-d$. The result of this test is shown in Fig. \ref{polsusc}b: for both marginal and standard model eq. \eqref{PPmagic} holds. We conclude that in the symmetry-broken phase, close to $T_c$, the marginal model has no difference from a standard $\mathrm O(n)$ ferromagnet, as it was expected from its symmetry properties.

\subsection{Low-temperature behaviour of the marginal model}

\begin{figure}[]
\centering
\begin{subfigure}{0.5\textwidth}
\includegraphics[scale=0.27]{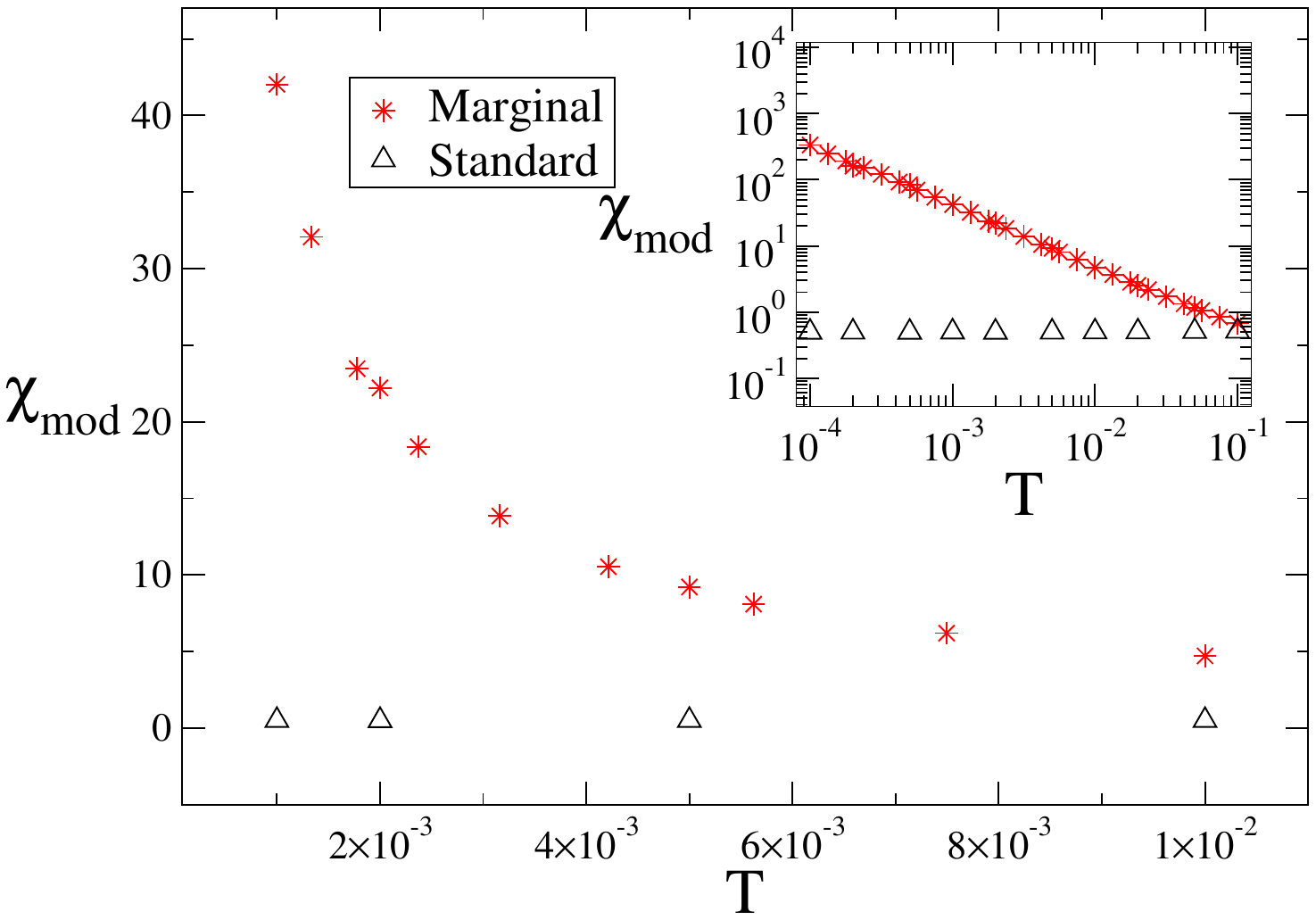}
\caption{}
\end{subfigure}%
\begin{subfigure}{0.5\textwidth}
\includegraphics[scale=0.27]{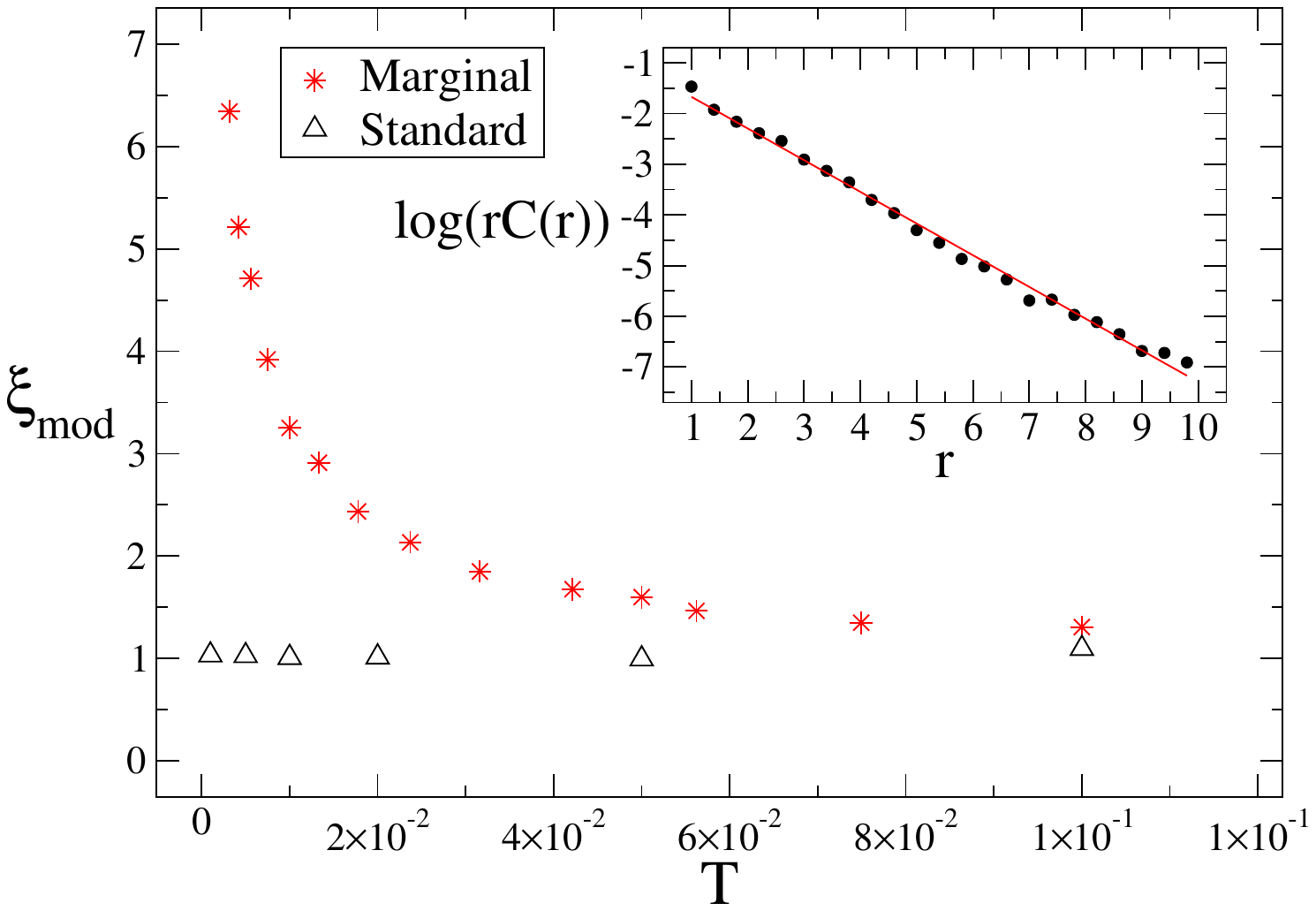}
\caption{}
\end{subfigure}
\caption{(a) Modulus susceptibility as a function of temperature. Simulations parameters: $L=20$, $h=0$. The modulus susceptibility of the marginal model grows for vanishing temperature while the standard model's remains small across the whole symmetry-broken phase. Inset: same plot in log-log scale. (b) Modulus correlation length as a function of temperature. Also this quantity grows for the marginal model as $T \to 0$, while it remains finite in the standard model. Inset: $\log (r\,C(r))$ vs $r$ in the marginal model, showing that the anomalous dimension is very small, so that a fit of $\xi$ is not problematic in this representation. $T=0.05$.}
\label{margpic}
\end{figure}

At the mean-field level, the interesting new feature of the marginal model, namely the growth of the modulus susceptibility,  emerges for small $T$; hence, this is the regime we need to explore with simulations. In complete analogy with the standard susceptibility, the modulus susceptibility is defined as the sum of the correlations of the fluctuations of the modulus,
namely, 
\begin{equation}
\chi_\mathrm{mod} 
= \frac{\beta}{N} \sum_{i,j} \langle\sigma_i \sigma_j \rangle-\langle\sigma_i\rangle\langle\sigma_j\rangle \label{chimod}
\end{equation}
where $\sigma_i\equiv|\boldsymbol\sigma_i|$ is the modulus of the microscopic degree of freedom. It is essential to understand that the longitudinal susceptibility, $\chi_\parallel$, calculated in the previous Section, is {\it not} the same as the modulus susceptibility, $\chi_\mathrm{mod}$ \cite{pata1973chi_longi}. Transverse and longitudinal modes are the projection of the fluctuations along the directions perpendicular and parallel to the spontaneous order parameter, and for this reasons they are both coupled with the phase fluctuation $\theta$: indeed, when $\theta=0$, not only there is no transverse fluctuation, but there is no longitudinal fluctuation either. Instead, a fluctuation of the modulus may occur also for $\theta=0$, because the modulus is the degree of freedom truly orthogonal to the phase. This is the reason why, in a standard $\mathrm O(n)$ ferromagnet, $\chi_\parallel$ diverges in the whole broken-symmetry phase (for $h=0$), while $\chi_\mathrm{mod}$ does not. The divergence of $\chi_\mathrm{mod}$ is the new physical feature we are after, and indeed Fig. \ref{margpic}a shows that our guess was correct: while $\chi_\mathrm{mod}$ of the standard model remains finite throughout the whole symmetry-broken phase, in the marginal model the modulus susceptibility {\it grows} quite rapidly approaching $T=0$. In order to calculate the precise exponent of this growth we need to perform finite-size-scaling analysis, and to do that, we
need a correlation length.

\begin{figure}[t]
\centering
\includegraphics[scale=0.3]{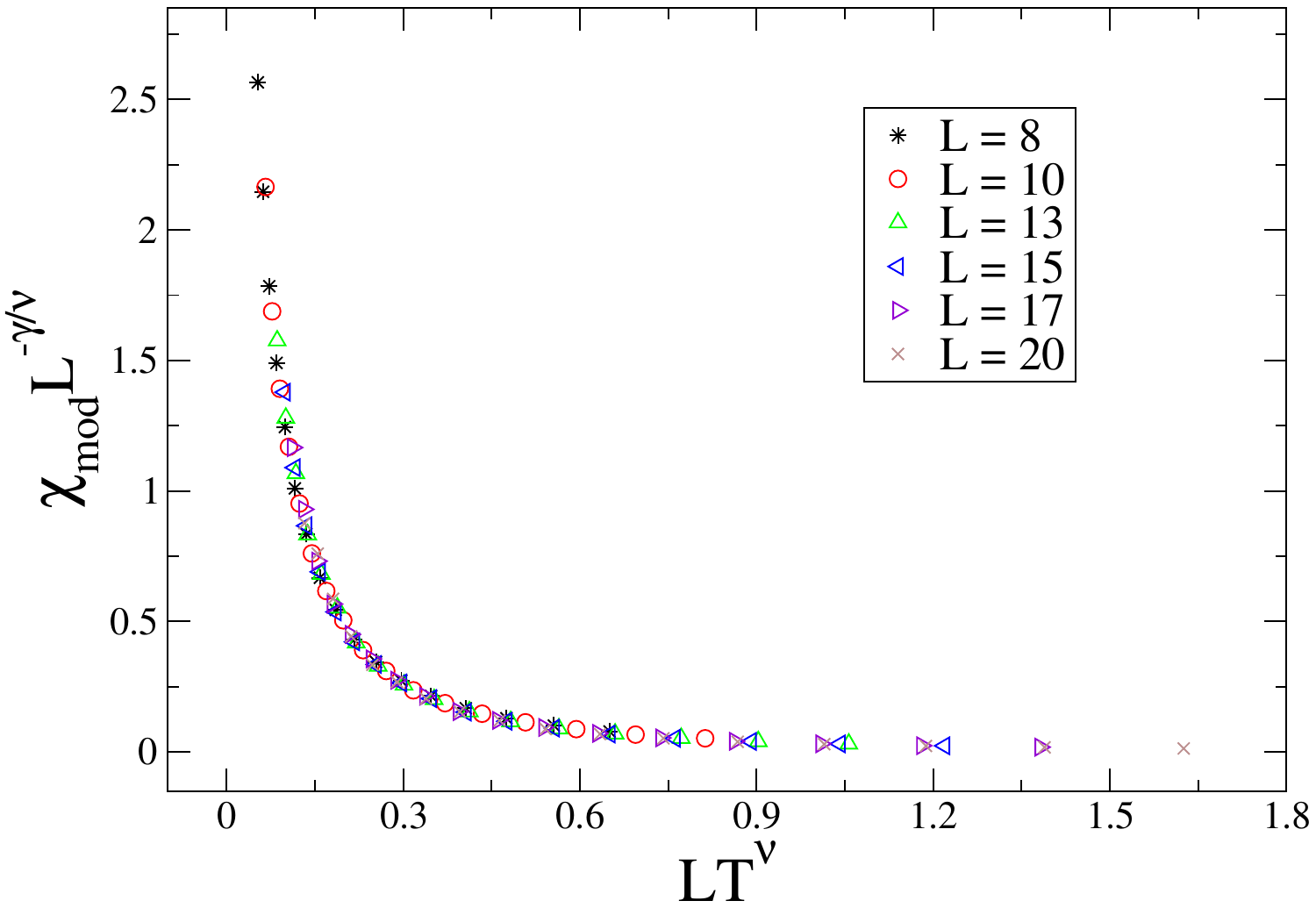}
\caption{Finite-size scaling (FSS) analysis of the marginal potential. Simulation parameters: $h=0$. Scaling of the modulus susceptibility for various low temperatures $T\in[10^{-4}:10^{-2}]$ and using all the simulated sizes of the system. The collapse of the curves for different sizes occurs with the critical exponents $\gamma = 1.08$ and $\nu = 0.55$. We tuned the critical exponents using a chi-by-eye procedure.}
\label{critexp}
\end{figure}

The normalized connected correlation function of the modulus fluctuations, $C_\mathrm{mod}(r)$, is defined in the standard way,
\begin{equation}
C_\mathrm{mod}(r)=C_\mathrm{mod}^u(r)/C_\mathrm{mod}^u(0)
 \quad , \quad 
C_\mathrm{mod}^u(r) =\frac{1}{q(r)} \sum\limits_{i,j}  \left(\langle\sigma_i \sigma_j \rangle-\langle\sigma_i\rangle\langle\sigma_j\rangle \right) \delta(r_{ij} - r)
\label{cfMC} 
\end{equation}
where $q(r)= \sum\limits_{i,j} \delta(r_{ij} - r)$ is the number of pairs at distance $r$ (because we are on a static lattice, such number does not fluctuate). Normalization is important, as we want to capture the change in the correlation only due to the change of its spatial span, that is of its correlation length, not of its amplitude. Following the general convention, we write the connected correlation function for large $r$ as,
\begin{equation}
C_\mathrm{mod}(r) \sim \frac{e^{-r/\xi}}{r^{d-2+\eta}}
\label{formcfMC}
\end{equation}
where $\xi$ is the correlation length. The anomalous dimension $\eta$ \cite{kardar_book} is usually much smaller than 1, and a semi-log plot of $r\,C_\mathrm{mod}(r)$ in $d=3$ (Fig. \ref{margpic}b, inset) shows that the marginal model close to $T=0$ is no exception; hence, for fitting purposes, we neglected $\eta$ and we work out $\xi$ from the linear fit of $\log(r\,C_\mathrm{mod}(r))$. The results are reported in Fig. \ref{margpic}b: a clear increase of the correlation length of the modulus is observed in the marginal model by lowering the temperature, while it is not observed in the standard model. This is interesting, as it means that {\it bona fide} long-range correlations (eventually scale-free for $\xi \gg L$) are truly developed by the new model in the deeply ordered phase, which is exactly what we were after in order to explain flock phenomenology. It therefore seems that we have indeed resolved the paradox: we have large correlations coexisting with large order parameter.

The sharp increase of both the susceptibility and the correlation length, supports the idea that in the marginal model $T=0$ acts as a true critical point. To test this hypothesis, we run finite-size-scaling (FSS) analysis \cite{privman1990finite, kadanoff1966introduction}. If in the bulk we have,
\begin{equation}
\chi_\mathrm{mod} \sim T^{-\gamma_\mathrm{mod} } \quad \quad , \quad \quad \xi_\mathrm{mod} \sim T^{-\nu_\mathrm{mod} }
\end{equation}
FSS prescribes that for finite $L$ one should observe the following  scaling form of the susceptibility \cite{privman1990finite},\
\begin{equation}
\chi_\mathrm{mod} = L^{\gamma_\mathrm{mod} /\nu_\mathrm{mod} } f\left(L/\xi_\mathrm{mod} \right)= 
L^{\gamma_\mathrm{mod} /\nu_\mathrm{mod} } f\left(LT^{\nu_\mathrm{mod}} \right) \ ,
\end{equation} 
where $f$ is an unknown scaling function. Therefore, we can determine the exponents $\gamma_\mathrm{mod}$ and $\nu_\mathrm{mod}$ by  plotting $\chi_\mathrm{mod}L^{-\gamma_\mathrm{mod}/\nu_\mathrm{mod}}$ vs $LT^{\nu_\mathrm{mod}}$ and imposing a collapse of all the data obtained at different sizes and temperatures. The results of FSS are reported in Fig. \ref{critexp}: we obtain a very good FSS collapse with the following critical exponents, 
\begin{equation}
\gamma_\mathrm{mod}  = 1.08 
\quad , \quad
\nu_\mathrm{mod}  = 0.55 
\end{equation}
Notice that the exponent $\gamma_\mathrm{mod}$ is comparable with its mean-field counterpart, and that both exponents are
quite close to their standard values at the finite $T_c$ \cite{ZJ1980critexp, Holm1993critexp}. Notice also that these exponents imply $\xi_\mathrm{mod} \sim \chi_\mathrm{mod}^{1/2}$, namely that the growth of $\xi_\mathrm{mod}$ is much slower than that of $\chi_\mathrm{mod}$, as it can be clearly appreciated from Fig. \ref{margpic}. The excellent FSS behaviour we find suggests that, for the marginal model, $T=0$ behaves indeed as a true critical point.

\section{Conclusions and perspectives}

At least at the theoretical level, the marginal model offers a successful way to resolve the apparent paradox of natural flocks: scale-free correlations of the modulus coexisting with strong collective order. The marginal model has a new critical point at $T=0$, where modulus susceptibility and correlation length diverge, so that strong correlations of the modulus are in fact not merely coexisting with, but a {\it consequence of} collective order. Moreover, all the phenomenology of the standard $\mathrm O(n)$ models, in particular the order-disorder transition and the Goldstone modes, is preserved. Of course, the phenomenology we found deserves a deeper theoretical study. The only theory we did here was mean-field, which is quite minimal. It would be important to write a Landau-Ginzburg-like field theory and investigate the zero-temperature critical point through the renormalization group. This is left for future work.

The zero-temperature critical point of the marginal model indicates that, provided that the system's order is strong enough, the critical point will give rise to large modulus correlation, eventually scale-free when $\xi_\mathrm{mod}$ becomes larger than $L$.  An interesting prediction of the marginal model is therefore that there should be a link between polarization and modulus correlation length. More precisely, the MC simulations give, 
\begin{equation}
\xi_\mathrm{mod} \sim \frac{1}{T^{1/2}} \sim \frac{1}{(1-m)^{1/2}} \ .
\label{pongo}
\end{equation}
It would be interesting to test this prediction in real data; however, if real flocks live in the deeply scale-free regime, $\xi_\mathrm{mod} \gg L$, it could be hard to test \eqref{pongo}, because one is unable to measure the bulk correlation length in the scale-free regime. Nevertheless, we hope that a more refined analysis of real flocks data may suggest some 
way out of this problem. Also, we note that in order to test the marginal potential on real flocks data, one should study the parameter space $(J,\lambda,T)$ carefully, so to match the experimental values of the polarization {\it and} of the modulus correlation length. This certainly requires further numerical and analytical work.

It is worth noticing that, at the biological level, a single particle marginal potential is nothing particularly strange: it simply means that a single bird has largish non-Gaussian fluctuations of the speed around its physiological value (about $10$ms$^{-1}$ for starlings). But because of the power $4$ in the marginal potential, these fluctuations are still very much bounded, a physiologically crucial requirement that was not met by the approach of \cite{bialek+al_14PNAS}. The interesting point suggested by the marginal model is that, even though single-bird fluctuations of the speed are non-Gaussian, once the bird interacts with other animals within the group, the effect of fluctuations is that the speed correlation length and susceptibility decrease, meaning that also individual speed fluctuations are {\it depressed} by the social interaction. We find this mechanism rather interesting, also at the biological level.

As we have already remarked in the Introduction, the marginal model shows that even in an equilibrium system it is possible to have scale-free behaviour of the modulus degree of freedom in a phase with high polarization. This, of course, does not demonstrate that the right explanation of that empirical phenomenon is an equilibrium (or quasi-equilibrium) one. But it does prove that intrinsic off-equilibrium explanations, from the self-propelled nature of active systems, are not strictly necessary.
Of course, the next step would be to study a self-propelled generalization of the marginal model, where the variables $\boldsymbol \sigma_i$ become real velocities, $\mathbf v_i$. This can be done merging the marginal model with preexisting Vicsek-like models of self-propelled particles \cite{vicsek+al_95, cavagna+al_15, toner+al_95}, through the equations,
\begin{align}
\frac{d\mathbf v_i}{d t} &= -\frac{dH(t)}{d\mathbf v_i} + \mathbf\zeta_i \ ,\\
\frac{d\mathbf x_i}{d t} &= \mathbf v_i  \ ,
\end{align}
where the ``Hamiltonian'' now is time dependent through the adjacency matrix, $n_{ij}(t)$ \cite{mora2016local} and the potential within $H$ is the marginal one. We would be very surprised if such a self-propelled model, in its symmetry-broken phase, did not develop the same scale-free modulus phenomenology as the equilibrium marginal model we studied here (much as the Gaussian model of \cite{bialek+al_14PNAS} works both at and off-equilibrium). Yet one should check.

We conclude with a general remark about correlations in collective biological systems. Speed fluctuations in flocks are not the only instance in which correlations are stronger than any purely mathematical reason would prescribe (as Goldstone's theorem). Neural assemblies \cite{Schneidman:2006p1273}, insect swarms \cite{kelley2013emergent}, bacterial clusters \cite{zhang2010collective}, and proteins \cite{tang2017critical} are examples of this phenomenon. In such cases, one way to explain the data is to look for a parameter $x$ that has been tuned to be close to some critical point, thus giving rise to large susceptibility and correlation length \cite{mora+al_11}. There are two tricky issues in this approach: first, in finite-size systems, the maximum of the correlation length occurs at a value of $x$ which depends on the system's size; hence, the system cannot be hard-wired once and for all at the bulk critical point. Secondly, what is delicate about `tuning' is that the parameter cannot be too small, nor too large; some precision is required, which may be hard to achieve, although most fascinating when it occurs \cite{Bialek1990retina}. The model we introduced here offers an interesting alternative to `tuning' at criticality. When the critical point is at the {\it boundary} of some parameter (the temperature, in the marginal model), first we do not have to worry about the finite-size dependence of the optimal value of the parameter, as this remains stuck at the boundary for all sizes; secondly, pushing the parameter as close as possible to its boundary value is somewhat less conceptually demanding than tuning it at its size-dependent optimal value. It may be that this kind of ``boundary critical point" has therefore some general relevance in collective biological systems with anomalously large correlations.

One could ask, however, whether we are merely shifting the problem from something quite tricky (tuning at criticality), to something else even more tricky (a marginal potential). In this respect, we first note that there is no obvious biophysical constraint dictating that individual fluctuations must be marginal: even though, of course, physiology requires most kind of biological fluctuations to be bounded (certainly speed fluctuations in birds must be!), we see no specific reasons why they should be bounded in a non-Gaussian way, i.e. with a power larger than $2$ in the logarithm of the probability distribution. This fact suggests two kind of considerations: first, one could invert the argument and claim that it is actually the Gaussian form to be rather specific, with all larger powers being more generic; this is not overly convincing, though, as the vanishing of the first nontrivial term in the Taylor expansion of any function seems to require some non-trivial justification. Second, we are left with the functional argument: the consequence of a marginal bounding potential is to have scale-free correlations of the fluctuations coexisting with large order parameter, which could be a biologically relevant condition for a collective system to work efficiently. If this is the case, individual marginal fluctuations may be the result of an evolutionary process, rather than an accidental biophysical constraint. However, for this interpretation to be more firmly established, one would need a solid proof that large groups that lack scale-free correlations {\it and} large order parameter, fail to achieve some important biological aim. Irrespective of how reasonable this seems to be, we believe that more work, theoretical, numerical and, most importantly, experimental, needs to be done to prove this point.

\section*{Acknowledgements}
\noindent
We thank Stefania Melillo and Massimiliano Viale for their help in developing and optimizing the code; ACu and LDC thank Giulia Pisegna and Federica Ferretti for many useful exchanges of views, ACa and ACu thank Lara Benfatto for her opinion and advice. We also thank William Bialek for interesting discussions and for several suggestions about the work. ACa., ACu. and LDC are supported by ERC Advanced Grant RG.BIO -- contract No. 785932. Corresponding Author: ACu.
\vskip 1 truecm 

\bibliography{general_cobbs_bibliography_file_2018_12}

\end{document}